# Emission of charged particles from laser-induced germanium ecton, vacuum spark and vacuum arc

V. Porshyn

*Faculty of Mathematics and Natural Sciences-Physics, University of Wuppertal, 42119 Wuppertal, Germany*

Electronic mail: porshyn@uni-wuppertal.de.

**Abstract**

The highly resolved temporal evolution of laser-induced micro-explosions on a germanium surface is studied in a triode configuration for various gate charge levels and cathode currents. Electron emission from individual spots is directly imaged with a luminescence screen, showing that the opening angle of the source is about 30°. Electron bunches of several nanocoulombs per pulse in a time interval of about 150 ns are directly extracted to the anode without vacuum breakdown in the cathodic gap. When breakdown occurs, a remarkable change in the arc behavior of a threshold gap potential of around 1 kV is observed, which hints at two different evaporation mechanisms that depend on the cathodic fall of an individual spot. Therefore, for voltages well above the threshold, a fast gate discharge is observed within the first 100–200 ns, followed by fundamental plasma oscillations and an electron emission of several μC per pulse from the plasma boundary. Additionally, highly efficient emission of germanium ion clusters occurs, evidencing a stable two-fold electron multiplication in the plasma, with a charge of several μC per pulse below the threshold.

---

# I. Introduction

Laser-induced surface excitation of solids with short pulses, i.e., well below several tens of nanoseconds, can introduce numerous interesting phenomena, ranging from the simple remelting of an upper surface layer of a solid to intense nuclear excitations, for laser fluences above ~$10^7$ W/cm$^2$ and above ~$10^{18}$ W/cm$^2$, respectively [1,2]. Small laser fluences are viable candidates in triggering micro-explosions, so called ectons, in the presence of electrostatic fields and are an efficient source of low-energy electrons and ions. After the initiation of the ecton the solid target acts as a cathode, which starts to emit electrons efficiently from overheated surface imperfections. This process leads to local explosions and the formation of craters, as described consistently by recent theories of vacuum breakdown under high electric fields, i.e., above approximately 1 MV/m [3,4]. Thereby, electron current densities from a single ecton up to ~$10^8$ A/cm$^2$ are possible. However, the total current of the ecton is limited due to its finite dimension of several tens of micrometres, and thus an arc discharge typically involves numerous spots to maintain itself [5,6]. Nevertheless, individual ecton sites are still interesting to researchers, since initial beam size and shape are important parameters in any practical application [7]. In addition, vacuum arc plasma typically consists of different species, e.g., micro droplets, multiple ionized ions, and neutrals, which should also be considered.

Based on previous work on germanium [8], the emission characteristics of laser-induced micro-explosions and the basic properties of ignited plasma are investigated here, with high temporal resolution using an oscilloscope in a compact triode configuration. The spatial distribution of the electron beam is imaged directly with a luminescence screen. An electrical circuit with a tuneable gate capacitance and cathodic resistance is used to suppress and analyze micro-explosions in their different evolutionary stages.

## II. Experimental technique

A triode configuration was assembled inside a vacuum tube (KF40 type), as shown in Fig. 1. A highly p-doped conductive germanium crystal (111), with an initially mirror-like flat surface was used as a cathodic target and placed in front of a copper gate with an aperture of 2 mm in diameter ($r_a$). The distance between the cathode and the gate ($d_{c-g}$) was fixed with a ceramic spacer to 1 mm. The anodes consisted of commercially available optically transparent conductive glass plate coated with an indium tin oxide (ITO) layer, mounted using ceramic screws above the gate. A glass plate coated with ZnO:Zn phosphor, which had a phosphor free region at its center to allow a laser beam to pass, was used to image incident electrons. The imaging was done with a green filter to avoid an overexposure due to laser illumination and plasma glow. The distance between the gate and the anode was 5 mm and 10 mm for the ITO layer with and without phosphor, respectively. Measurements were carried out at a base vacuum level of around $10^{-6}$ Pa. However, vacuum levels of up to 3 Pa did not significantly affect emission characteristics. All electrodes were analyzed using common scanning electron microscopy and energy dispersive x-ray techniques before and after the measurements.

The illumination source was the third harmonic of a pulsed Nd:YAG laser with a wavelength of 355 nm and pulse duration of 3.5 ns. The beam was focused with a plano-convex lens to a circular spot with a $1/e^2$ beam size of 100 µm; a peak intensity on the order of $10^7$ W/cm$^2$ was applied to initiate ectons.

A fast voltage response with negligible line reflections was secured with compact connection lines for the cathode and the anode, with lengths under 5 cm to the ground. The total parasitic capacitances of the anode and the cathode were measured to be below 2.4 pF and 6 pF, respectively. In regular configuration, the voltage response was acquired as a voltage drop at serial resistances ($R_c$) and ($R_a$) by an oscilloscope (Infiniium 54852A) with a 2-GHz bandwidth. The input of the oscilloscope was upgraded by high-frequency-capable attenuators (A) with input impedances of 100 kΩ and 50 Ω for the cathode and the anode, respectively. The gate was charged via a 10-MΩ resistor to a maximum gate voltage ($U_{g,max}$) using a high voltage power supply (FuG HCN 35-35000). The minimum initial gate capacitance ($C_g$) was found to be 22.1 pF.

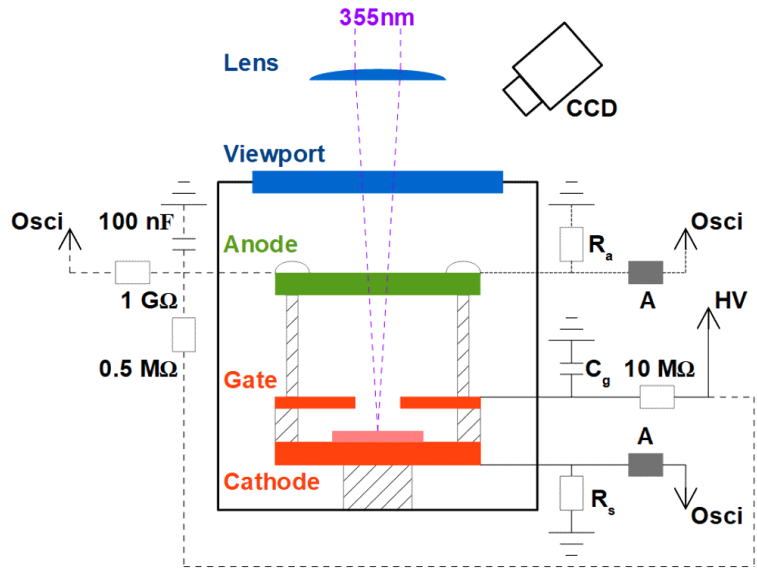

**FIG. 1.** Scheme of the triode setup with the electrical circuit diagram. The main and alternative connections are highlighted with solid and dashed lines, respectively.

In addition, the circuit was modified for beam imaging to accelerate the extracted electrons above 1 kV, and thus to provide a measurable luminescence. For this purpose, the gate and the anode were coupled with a 0.5-MΩ resistor. The anodic response was roughly monitored by a low bandwidth network, and an additional 100-nF capacitor was used to stabilize the anodic voltage.

## III. Results and discussion

### A. Electron emission from laser-induced ectons and low-density vacuum sparks

The emission of free electrons from a surface plasma for a fixed geometry and gap voltage depends primarily on the initial gate charge and the magnitude of the discharge current. As already mentioned, both these parameters can be controlled by the series resistance of the cathode and the capacitance of the gate. For sufficiently small charge values of several hundred nanocoulombs, a highly dense plasma is present at an ignition point and electron emission results mostly from a laser-induced ecton, with a typical radius ($r_c$) of around 10 µm, as previously reported for an unlimited current circuit [8]. By limiting the amount of current flowing throw the cathode, small craters were created just as expected [9]. More importantly, the lifetime of a laser-induced ecton, unlike the lifetime of ectons induced by a high voltage pulse, was limited to (4.1±0.4) ns [10], see Fig. 2. Obviously, in this case the explosion process was correlated with the pulse duration of the laser beam. In this short time period, the emitted electrons travelled significantly faster than evaporated neural atoms towards the electrodes. Furthermore, the voltage response showed a systematic delay of around 0.67 ns between the anodic peak voltage and the cathodic peak voltage, which corresponds quite well with the expected electron time-of-flight ($t_{c-a}$) for the cathode-anode distance ($d_{c-a}$) of 11 mm and the initial gate voltage of 3 kV. Therefore, the shortest time-of-flight for non-relativistic electrons, assuming a constant electron velocity, can be estimated as follows:

$$t_{c-a} = d_{c-a} \sqrt{\frac{m_e}{2eU_{g,max}}} \tag{1}$$

with elementary charge e and electron mass $m_e$. A typical gate transmission ($T_g$) for a spot, located at the center of the gate aperture, was measured to be at least 10% per ecton. Thus, a proportion of electrons was attracted to the gate, subsequently reducing the initial acceleration in the cathodic gap. The rising cathodic and anodic voltage responses introduce an additional decrease in acceleration, leading to a broadening of the anodic response. However, the rest gate voltage after termination of the ecton was found to be above 80 %. Hence, the electron time-of-flight can be fairly determined with an uncertainty of about 100 ps.

After termination of the ecton due to phase-explosion phenomena, a further relaxation of the excited plasma occurs [11–13]. The ionized and neutral evaporated material continues to expand adiabatically, i.e., without heat transfer to the environment, and transforms into a vacuum spark. As measured, the spark also supports an emission of free electrons, as long as the ignition point remains active. The duration of this emission corresponds to the lower tail of the voltage response in Fig. 2.

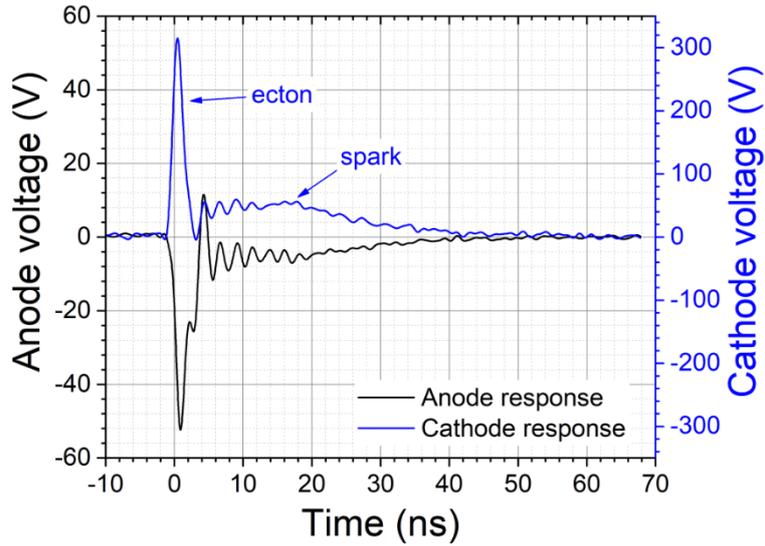

**FIG. 2.** Time-resolved voltage response with a fingerprint of the laser-induced ecton in the first few nanoseconds during a single-pulse laser illumination for $C_g$ = 22.1 pF, $R_c$ = $R_a$ = 50 Ω, $U_{g,max}$ = 3 kV.

Laser-induced vacuum sparks can also be initiated without an extensive local explosion, see Fig. 3(a). This type of emission appeared to be dominant, in particular, for more eroded areas, i.e., after a large number of pulses. A typical pulse shape of a spark can be divided into at least two different time periods. The first, of about 50 ns, is related to the time-of-flight of neutral, single, and multiply charged particles to the gate and typically shows a pronounced current increase up to the point where the interaction between the gate and the expanded germanium plume becomes significant, i.e., "gap bridging" in Fig. 3. The mean velocity of the neutral particles is roughly estimated from the data to be approx. 2 x $10^4$ m/s, in good agreement with the literature [14]. The following time period can be characterized by variable voltage behavior, which is directly regulated by the external circuit. In particular, for sufficiently high $R_c$ values, the current is limited and avalanches in the cathodic gap are suppressed. A similar process was previously observed for the controlled explosions of liquid metal tips, with a size of several micrometers [15]. In total, charge values of several nC per pulse was extracted to the anode without a complete breakdown in the cathodic gap. The corresponding gate transmission efficiency in Fig. 3 reached up to 6.5 %. The lifetime of vacuum sparks was primary limited by the discharge of the gate below 800 V, as the macroscopic electrical field decreased below 1 MV/m. The discharge cut-off can be explained in this case by a rapidly decreasing field emission from the ignition point, for a decreasing electrical field strength in the cathodic gap.

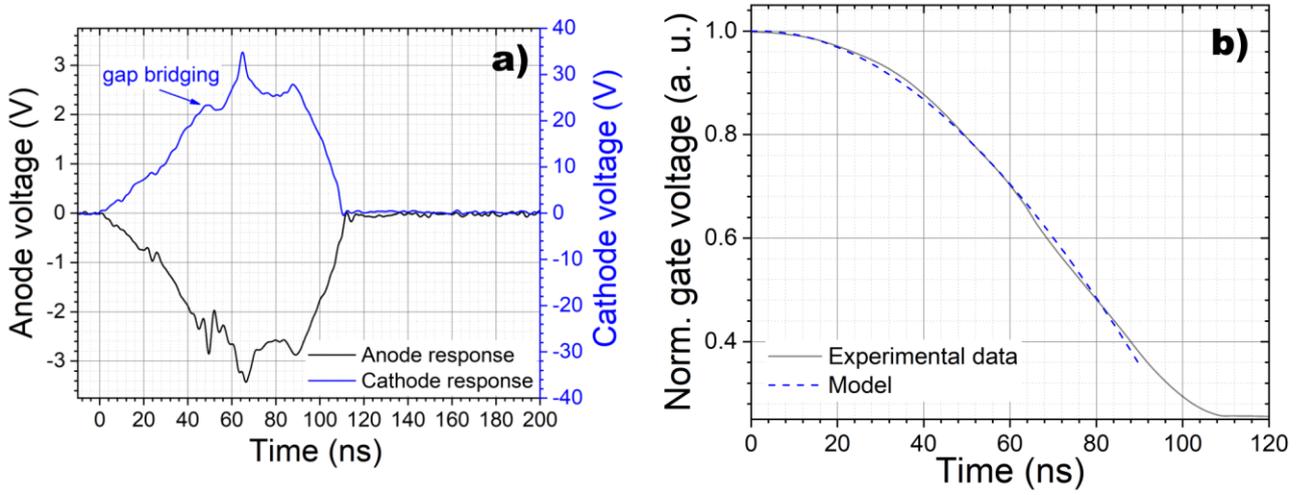

**FIG. 3.** (a) Typical time-resolved voltage response of a vacuum spark induced by single-pulse laser illumination for $C_g = 22.1$ pF, $R_c = R_a = 50$ Ω, and $U_{g,max} = 3$ kV. (b) Normalized gate voltage response as obtained from the data in (a) and the corresponding fitting of the data up to 90 ns, with $\alpha = 2.17$.

Assuming all electrons move free and collisionless to the gate, the equation for the time-dependent cathodic current $I_c$ can be derived by combining the following basic relations for the current, $I_c(t) = -en_e v_e(t) S$, and the acting force, $F(t) = m_e v_e(t)/t = -eE(t)$, where $n_e$ is the electron density, $v_e$ the electron speed, E the electric field, and S the cross-sectional area. The resulting relation becomes $I_c(t) = \alpha E(t) t$, with the proportional constant $\alpha = n_e e^2 S/m_e$. Moreover, the gate voltage follows $U_g(t) = U_{g,max} - I_c(t)t/C_g$ rather than regular capacitor discharge behavior. The time-dependent voltage drop of the gate can be estimated with the following expression:

$$U_g(t) = U_{g,max}\left(1 - \frac{t}{1 + R_c C_g + \frac{d_{c-g} C_g}{\alpha t}}\right) \quad (2)$$

which allows one to derive the electron density of the spark directly from the voltage response, as presented in Fig. 3(b). The estimated proportional constant was equal to 2.17; hence, the electron density at the ignition point, with crater radii of 1–10 µm follows to be of the order of $10^{17}$–$10^{19}$ m$^{-3}$. The resulting value appears to be significantly smaller in comparison with explosive electron emission initiated by field emission only, where a typical order of the electron density at the cathode is around $10^{27}$ m$^{-3}$ [16,17], and the discharge process is primary determined by the self-sustained lifetime of an individual ecton. Furthermore, the electron density is related to the Debye length $\lambda_D = (\varepsilon_o T_e/(n_e e))^{1/2}$, where $\varepsilon_o$ is the vacuum permittivity and $T_e$ the electron temperature in eV [18]. The Debye length of the laser-induced sparks follows to be in the same range as the size of the craters 1–10 µm.

## B. Charged-particle emission from laser-induced vacuum arcs

### *1. High current discharge*

For sufficiently high $C_g$ and low $R_c$ values the electron and neutral-atom density in the cathodic gap increases rapidly up to the breakdown down event; i.e., a collective ion motion in the gap becomes significant, and a plasma sheath forms close to the gate electrode. Occasional plasma arcs are initiated above 250 pF, and for values above 1 nF, the discharge becomes regular, with its specific voltage response and emission characteristics shown in Fig. 4. In this case, glowing of the cathodic gap was clearly visible with the naked eye. The voltage response oscillates in the breakdown stage. Oscillations are a fingerprint of the different plasma processes and are unique for a given cathodic material [19].

A typical discharge process starts with an emission current of around 100 A from a laser-induced ecton. The arc current begins to increase rapidly after the time interval of plume expansion, in which the evolution of the cathodic spot dominate the voltage response. Some of the observed high-frequency oscillations, with typical periods of 5–10 ns, probably originate from the interaction between the intense electron jets and the expanding plasma vacuum boundary. This plasma instability is predicted to occur for high current-density rise rates, around $10^9$ A/cm$^2$/ns, by kinetic current numerical simulations of plasma near the cathodic spots [20]. The maximum current was reached quite reproducibly for different gate capacities within the next 40±3 ns. This time frame is also known as the "commutation time" $t_c$ of the spark phase [4]. The current peak value at the end of the spark phase was derived from the linear trend of $R_c$ vs. $U_c$ for $R_c$ < 5 Ω to be 195±2 A for $C_g$ = 5.08 nF. In the next 35 ns the gate capacitor was completely discharged. However, part of the energy appeared to be stored in the magnetic field of the arc current. Hence, the plasma was perturbed and started to oscillate with a frequency ($f_p$), somewhere between its own ion-sound and fundamental ion plasma frequency [21,22]. If we take $f_p \approx 8.98Z(n_e)^{1/2}(m_e/m_i)^{1/2}$ as an upper limit for the frequency with Z = 1.4 as the charge state, the ratio $(m_e/m_i)^{1/2}$ = 1/365 as the electron-to-ion mass ratio, and an oscillation period of 400 ns, then the electron density of the arc in Fig. 4 can be calculated to be at least 5.27 x $10^{15}$ m$^{-3}$, close to that of previous arc studies for similar current values [23,24].

Remarkably, the anodic voltage response did not reflect any meaningful extraction of electrons from the plasma after gap bridging during the time intervals where the electrons were attracted to the gate. This result clearly demonstrates a field-shielding effect at this stage due to a plasma sheath [25], which forms in the aperture of the gate electrode and significantly decelerates electrons. In contrast, electrons were emitted from the open plasma boundary during the negative charge cycles of the gate. A total negative charge of several µC can be extracted to the anode in this way. Ion emission during the positive charge cycle appeared to be rather suppressed, probably due to a low ion density in the gate region.

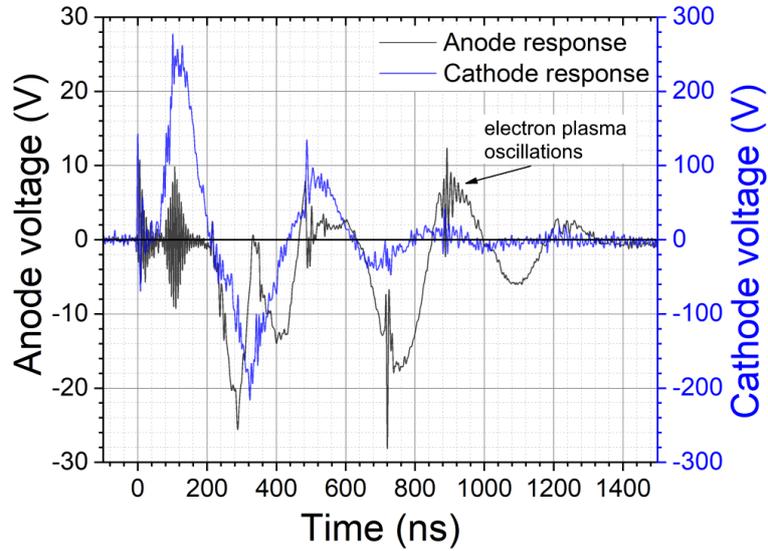

**FIG. 4.** Typical time-resolved voltage response for a vacuum arc induced by single-pulse laser illumination for $C_g = 5.08$ nF, $R_c = 1$ Ω, $R_a = 0.5$ Ω, and $U_{g,max} = 3.5$ kV.

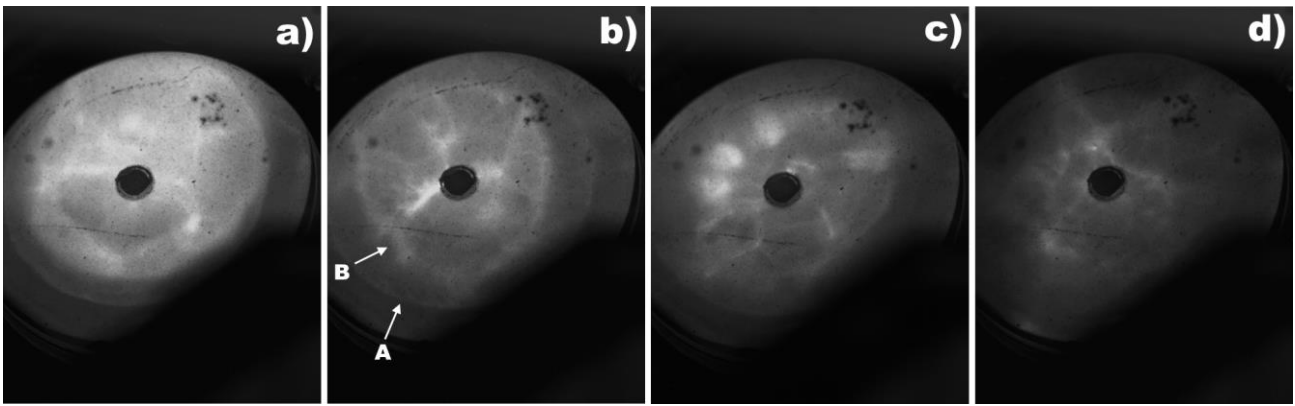

**FIG. 5.** Vacuum arc patterns obtained for $C_g = 2.24$ nC, $U_{g,max} = 3$ kV, and $R_c$ values of 1 Ω in (a), 4.7 Ω in (b), 10 Ω in (c), and 14.7 Ω in (d). The projected boundaries A and B, highlighted with the arrows, are related to the different origins of the emitted electrons. The real diameter of the dark area in the center is 2 mm.

Additionally, erratic polarity switching in the cathodic gap gave rise to electron plasma oscillations with resonance frequencies ($f_e$) of several hundred MHz, which introduced further voltage spikes in each oscillation cycle of cathodic and anodic voltage responses. These oscillations are related to the electron density as $f_e \approx 8.98(n_e)^{1/2}$ and can be used for its rough estimation [26]. Electron densities of around $10^{14}$–$10^{15}$ m$^{-3}$ follow for typical spikes with oscillation periods of 2–10 ns.

Figure 5 shows an enlarged projection of the cathodic spot, revealing a complex structure of individual arc events. A better focused emission from a laser-induced ecton overlaps on the screen with a diffuse electron emission caused by plasma instabilities and oscillations. Notably, during gap bridging, electrons are exracted directly to the anode, but the anodic potential becomes screened once the plasma sheath forms. Typically, two kind of reproducible boundaries are derived from the electron emission pattern, which are indicated in Fig. 5(b) as boundary A and boundary B. Boundary A always appears circular and is related to the gate aperture. The shape and position of boundary B are variable in correspondence with the individual emission sites on the cathode. From the geometry of the triode and typical crater dimensions, opening angles for boundaries A and B of 53°±2° and 30±2° are determined, respectively. For high emission currents, i.e., small $R_c$ values, the intensity of the pattern increases. Sufficiently detailed emission patterns were observed, in particular, for arcs in the current range of 50–100 A. The inhomogeneous structure of the pattern confirms that electrons were not uniformly emitted from the ignition point, which probably became the origin of numerous avalanches growing in size with increasing peak current. A local surplus of electrons on the screen is partly associated with branching electric discharge, i.e., Lichtenberg figure [27]. This demonstrates the presence of ionized channels extending transversely to the external electrical field, with an electron flow across the channel. The pattern becomes more point-like, the number of spots decreases, and the crater edges disappear with decreasing peak current. This behavior occurs because field electron emission from transient nanometer protrusions and outstanding edges is the major electron source in a vacuum arc [28,29].

## *2. Low current discharge*

Efficient emission of ionized germanium clusters, without an extensive visible glow in the cathodic gap, was measured for the current-limited circuit, as shown in Fig. 6(a). At the beginning of discharge in the first one hundred nanoseconds, the voltage characteristic was similar to that of previously discussed behavior. After the commutation time interval, the cathodic current increased to a peak value of 10.3 A. A further discharge was maintained by the continuous evaporation and explosion of random surface protrusions with a low field emission threshold, as already mentioned. Systematic voltage oscillations in the anodic response is strong evidence that the emission sites were unstable during the discharge process. The total duration of the process was limited to below 1.5 µs due to the decrease of the cathodic gap potential to below several hundred volts for a wide range of $R_c$, $C_g$, and $U_{g,max}$ settings. The time-of-flight of the ionized species from the gate to the anode could be directly extracted from the delayed positive anodic response to be mainly around 650 ns, as shown in Fig. 6(b). This time interval was found to be significantly long, as expected for a single germanium ion, even when considering the voltage drop of the gate. It can be concluded that clusters with at least 8–10 atoms were primarily extracted to the anode, in agreement with the fact that these types of clusters are most likely to be formed by germanium [30,31]. In particular, massive ionized clusters can readily overcome the potential difference of the cathodic gap, which was measured to reach an equilibrium voltage of around 500 V, and escape the plasma through the aperture. Their initial kinetic energy might be on the order of several hundred eVs and at the same time they are largely not scattered in the direction transverse to the microscopic electrical field. Furthermore, this ion current is expected to be below the saturation ion current of the arc $I_i = 2Zen_i(eT_e/m_i)^{1/2}M\pi r_a^2$, with Mach number M and ion density $n_i$ [32]. Hence, the ion density of the bunch of the ion clusters should be of the order of $10^{22}$ m$^{-3}$, assuming the following values: $I_i$ = 10 A, $T_e$ = 0.5 eV, Z = 1, M = 4.8, and $m_i$ = 654$m_p$.

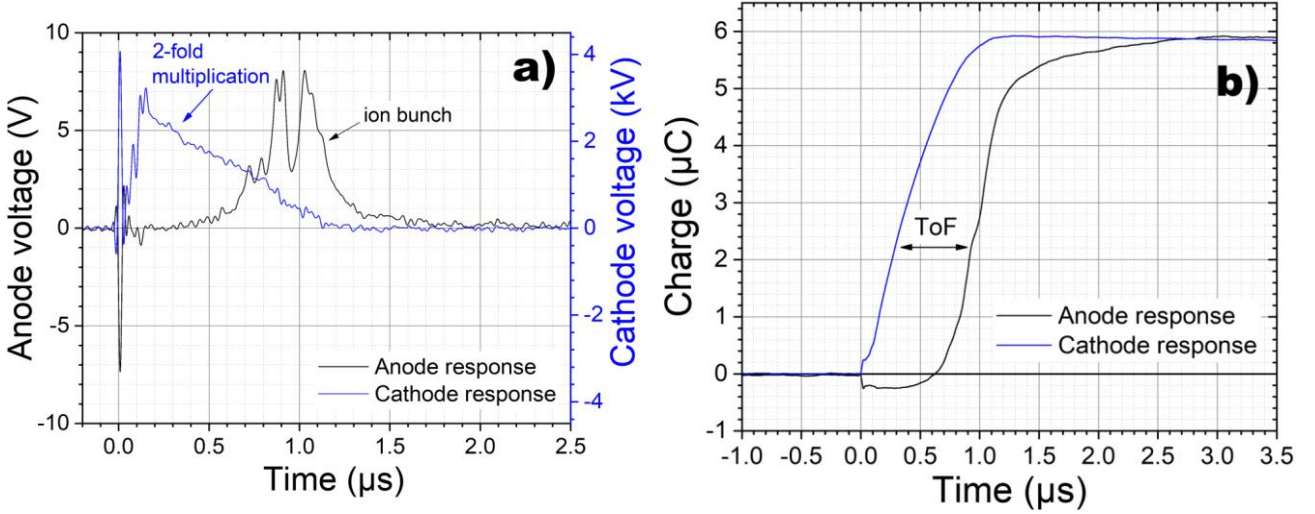

**FIG. 6.** Typical time-resolved voltage response in (a) and the corresponding charge behavior for a vacuum arc in (b) induced by single-pulse laser illumination, with $C_g$ = 3.46 nF, $R_c$ = 280 Ω, and $R_a$ = 0.5 Ω for $U_{g,max}$ = 3.5 kV. The time delay (time-of-flight, ToF) is related to the time-of-flight of the ion clusters.

## *3. Total emission characteristics*

The emission behavior of the vacuum arc is summarized in Fig. 7 for various $C_g$ and $R_c$ values and a fixed gate voltage. The most significant transition in emission behavior occurs for a cathodic resistance of around 10 Ω, as the gap potential drops to a critical value ($U_{crit}$), close to 1 kV, within the commutation time. This value is not fixed because of charge fluctuations, which are related to the ecton and can be roughly estimated from the temporal cathodic voltage response, with $U_{crit} \approx U_{g,max} - I_{c,max}R_c - (I_{c,max}t_c/(2C_g))$, e.g., U = 1128 kV for $I_{c,max}$ = 150 A, $R_c$ = 10 Ω, and $R_c$ = 3.44 nF. It can be assumed that the transition in emission behavior strongly depends on how the liquid surface can be destroyed by electrostatic forces, expressed as $U_{crit}/d_{c-g} \approx (2\gamma/(\varepsilon_0 r_c))^{1/2}$, where γ is the coefficient of surface tension [33]. After vacuum breakdown, i.e., when $d_{c-g}$ can be replaced by $r_c$, the corresponding transition potential becomes $U_{crit} \approx (2\gamma r_c/(\varepsilon_o))^{1/2}$; for instance, $U_{crit}$ = 1164 V for $\gamma$ = 0.6 N/m and $r_c$ = 10 µm for germanium [34]. Thus, evaporation from a "hot" crater becomes less destructive and more localized inside the crater, see Fig. 5(d). Following this trend, the Debye length at the cathode decreases, and the plasma electron density increases. As a consequence, microsecond, short, high-density, ion-cluster bunches with ion erosion rates of around 6.5 mg/C are efficiently extracted from the plasma with reduced cathodic currents of about 10 A, see Fig. 6. The magnitude of the erosion rate is sufficiently high in comparison with the values normally obtained by long-term vacuum cathodic arcs [35].

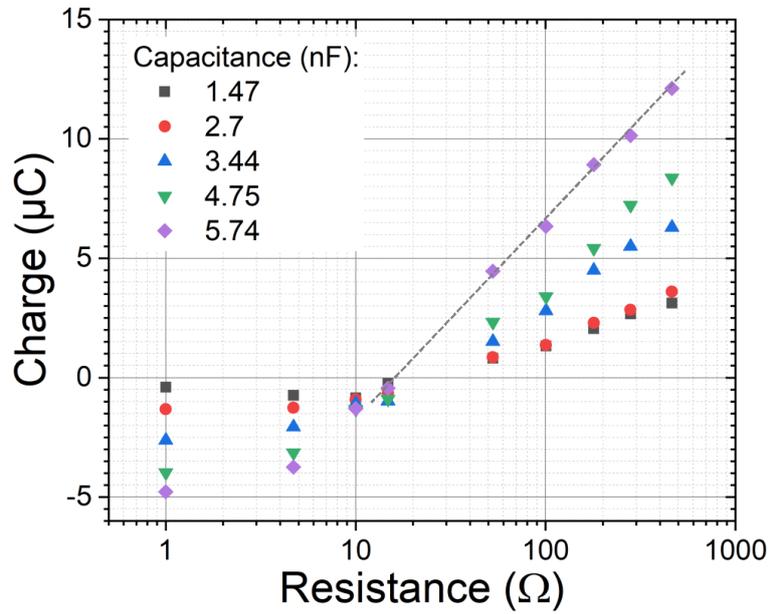

**FIG. 7.** Anodic charge (Q) vs. cathodic resistance ($R_c$) for various gate capacitances and a fixed gate voltage of 3.5 kV. The relative error of Q is estimated to be 5 %. The linear trend Q ~ ln($R_c$) is highlighted by the dashed line.

Finally, the anodic charge (Q) for $R_c$ values above the measured self-sustaining resistance of 10 Ω are adequately approximated by the relation Q ~ ln($R_c$), i.e., Q ~ ln($I_c$) is independent from the gate capacitance. The physical reason for this behavior has to do with the ionization and dissociation of neutral clusters [25,36]. The ionization rate increased for high $R_c$ values due to decreasing surface evaporation, and consequently, a low density of neutral evaporated gas. The cathodic current became sensitively dependent on the total ionization probability over the capacitor discharge time in this case. Indeed, the anode-to-cathode charge ratio exceeded values of around 1, see Fig. 6(b), evidencing a two-fold electron multiplication process; i.e., each electron arriving from the cathode ionized one cluster. Taking into account the regular discharge behavior of the gate capacitor $U_g(t) \sim \exp(-t/(R_pC_g))$ and, for simplicity, the Spitzer formalism $R_p \sim 0.53 \times 10^{-4} Z/(T_e)^{3/2} \ln\Lambda$ (with the Coulomb logarithm $\ln\Lambda$ [37,38]), $U_g(t) \sim \exp(-t/\ln\Lambda)$ is a function for which, in a high-density plasma, the Coulomb logarithm tends to zero and can be expanded around this point. Thus, the discharge voltage becomes a function of electron density, i.e., $U_g(n_e) \sim t_{dis}/n_e$ for a fixed time period ($t_{dis}$). The voltage integral over the electron density gives the following result: $U_{g,max} \sim \ln(n_e/n_{o,e})$, with $U_{g,max} \sim Q$ and $n_e \sim I_c$, in correspondence with the observed discharge behavior.

## IV. Conclusions

Micro-explosions on a germanium surface were initiated with a pulsed nanosecond laser under electrical fields of several MV/m. The magnitude of the explosion was regulated by the external circuit. Electron emission dominated if vacuum breakdown was suppressed; emission occurred from a more intense short-time laser-induced ecton and a subsequent vacuum spark, with transmission efficiencies of 10% and 6.5 %, through the gate aperture of the triode. The electron density of the vacuum spark was estimated to be of the order of $10^{17}$ m$^{-3}$ at the ignition point. Furthermore, the electron emission from the ignition point was directly imaged with a luminescent screen. The emission pattern revealed an inhomogeneous spatial distribution of the electrons, which could be related to their possible origin. The opening angle of the emission's active surface area was estimated to be around 30°.

A gate charge of several microcoulombs was necessary to complete vacuum breakdown in the cathodic gap and to create an arc. The emission characteristics of the vacuum arc were found to be strongly dependent on the rest potential in the cathodic gap after gap bridging occurred, for an introduced fixed gap size. Thus, a remarkable transition between two arc regimes, the high and the low current regime, occurred for a rest potential of approximately 1.1 kV. Arcs for rest potential values of well above 1.1 kV led systematically to a rapid gate discharge within the first 150 ns and characteristic plasma oscillations. The plasma electron density of the high-current regime was estimated to be of the order of $10^{15}$ m$^{-3}$. Here, intense electron emission from the arc to the anode was observed for only a negative half-cycle of an oscillation, i.e., a negative gate polarity. In contrast, positive ion clusters were mostly emitted from the arc in the low current regime. The discharge time interval and the electron density increased to 1.5 µs and $10^{22}$ m$^{-3}$, respectively. The emission process proved highly efficient; thus, up to 50% of the total capacitor charge could be extracted in the form of positive ion clusters. The introduced triode behavior is likely primarily related to the basic properties of germanium.

## Acknowledgment

The author would like to thank the German Ministry of Education and Research (BMBF) for financial support.